\documentclass[aps,prl,reprint,superscriptaddress,floatfix]{revtex4-2}

\usepackage[utf8]{inputenc}       
\usepackage{graphicx}
\usepackage{siunitx}              
\usepackage[T1]{fontenc}          
\usepackage{lmodern}              
\usepackage{amsmath}              
\usepackage{bm}                   
\usepackage{hyperref}             
\usepackage[mathscr]{eucal}       
\usepackage[normalem]{ulem}

\hypersetup{hidelinks, colorlinks=false}

\newcommand{\mf}{m_\text{F}}
\newcommand{\abs}[1]{\left\vert#1\right\vert}
\newcommand{\phil}{\varphi_\text{L}}

\makeatletter
\let\cat@comma@active\@empty
\makeatother

\begin{document}

  \title{Observation of two non-thermal fixed points for the same microscopic symmetry}
  
  \author{Stefan Lannig}
  \email{ntfp-basins@matterwave.de}
  \author{Maximilian Pr\"ufer}
    \thanks{current address: Vienna Center for Quantum Science and Technology, Technische Universit\"at Wien, Atominstitut, Vienna, Austria}
  \author{Yannick Deller}
  \author{Ido~Siovitz}
  \author{Jan~Dreher}
  \author{Thomas Gasenzer}
  \author{Helmut Strobel}
  \author{Markus K. Oberthaler}
  \affiliation{Kirchhoff-Institut f\"ur Physik, Universit\"at Heidelberg, Im Neuenheimer Feld 227, 69120 Heidelberg, Germany}
  
  \date{\today}
  
  \begin{abstract}
    Close to equilibrium, the underlying symmetries of a system determine its possible universal behavior.
    Far from equilibrium, however, different universal phenomena associated with the existence of multiple non-thermal fixed points can be realized for given microscopic symmetries.
    Here, we study this phenomenon using a quasi-one-dimensional spinor Bose-Einstein condensate.
    We prepare two different initial conditions and observe two distinct universal scaling dynamics with different exponents.
    Measurements of the complex-valued order parameter with spatial resolution allow us to characterize the phase-amplitude excitations for the two scenarios.
    Our study provides new insights into the phenomenon of universal dynamics far from equilibrium and opens a path towards mapping out the associated basins of non-thermal fixed points.
  \end{abstract}
  \maketitle
  
  Universality is a powerful concept for characterizing systems by means of effective models based on features that are independent of microscopic details. This concept led to the identification of universality classes for systems at and near equilibrium \cite{Hohenberg1977,Bray1994}. For example, close to a phase transition, they are characterized by a few universal exponents, which describe the relevant properties as functions of the distance to the critical point \cite{ZinnJustin2004a}.

  In such close-to-equilibrium scenarios, universal exponents are typically associated with the symmetry properties of the underlying Hamiltonian in the respective phase; generically, that means that for a fixed Hamiltonian, there exists a unique set of universal exponents and scaling functions. The situation changes for systems far from equilibrium. Here, possible excitations are richer and not strictly tied to the symmetry properties associated with the ground state of the Hamiltonian.
  Thus, universal phenomena far from equilibrium are expected to have more defining elements than the underlying microscopic Hamiltonian.

Motivated by renormalization-group ideas, the concept of non-thermal fixed points (NTFPs) has been developed as a means of characterizing universal types of dynamics far from equilibrium \cite{Berges2008,Nowak2010}. NTFPs imply universal dynamics, that is, time evolution which can be described by rescaling in time and space according to a set of universal exponents \cite{Berges2015,orioli_universal_2015}.
Quantum simulators utilizing ultracold atoms have proven to be a versatile platform for studying non-equilibrium behavior of quantum many-body systems \cite{polkovnikov_colloquium_2011,langen_prethermalization_2016,ueda2020quantum}. 
During recent years, studies of universal phenomena far from equilibrium have intensified, both in theory 
\cite{%
Berges2008,
Nowak2010,
Schole:2012kt,
Nowak:2012gd,
Berges2014_attractive_basin,
Hofmann2014a,
Berges2015,
Maraga2015a.PhysRevE.92.042151,
orioli_universal_2015,
Williamson2016a.PhysRevLett.116.025301,
Bourges2016a.arXiv161108922B.PhysRevA.95.023616,
Karl2016,
Fujimoto2018a,
Walz:2017ffj.PhysRevD.97.116011,
Mikheev2019_leeft,
schmied_non-thermal_2019,
Fujimoto2018b.PhysRevLett.122.173001,
Williamson2019a.ScPP7.29,
Schmied:2019abm,
Gao2020a.PhysRevLett.124.040403,
Wheeler2021a.EPL135.30004,
Gresista:2021qqa,
RodriguezNieva2021a.arXiv210600023R,
Preis2023a.PhysRevLett.130.031602}
and experiment \cite{%
prufer_observation_2018,
erne_universal_2018,
Johnstone2019a.Science.364.1267,
Navon2018a.Science.366.382,
glidden_bidirectional_2021,
GarciaOrozco2021a.PhysRevA.106.023314,
Huh2023_2universalities,
Martirosyan2023mml}.
Universal scaling associated with NTFPs was observed experimentally in different systems \cite{prufer_observation_2018,erne_universal_2018,Johnstone2019a.Science.364.1267,glidden_bidirectional_2021}; recently, distinct universal phenomena, depending on the Hamiltonian symmetries, have been reported \cite{Huh2023_2universalities}.

\begin{figure} 
    \centering
    \includegraphics[width=0.9\linewidth]{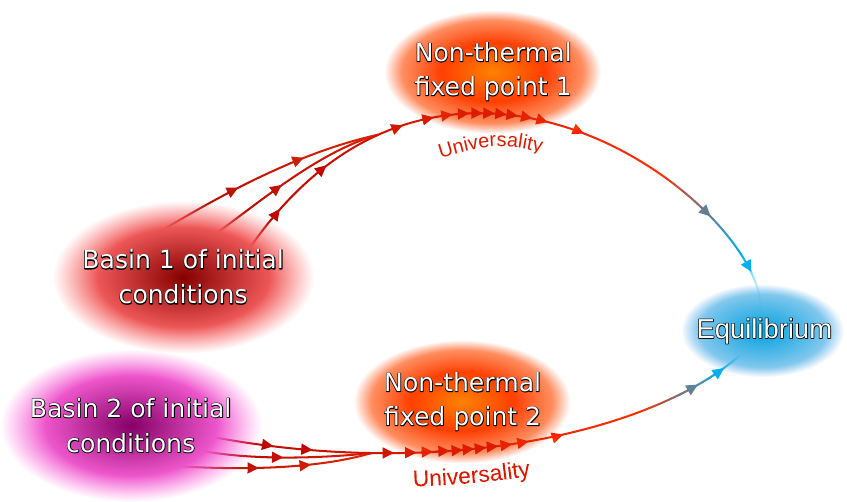}
    \caption{\label{fig0} 
    Schematic representation of the equilibration of a many-body system starting from different initial conditions. 
    During this evolution a transient period of slow self-similar behavior in the vicinity of a NTFP may occur.
    The associated basins are defined by the set of initial conditions which evolve towards the same NTFP.
    This is the simplest scenario, more complicated ones are possible, such as the existence of associated basins that lead to a subsequent time evolution to two or more NTFPs.}
  \end{figure}

Far from equilibrium, there is numerical evidence that the universal behavior can exhibit a dependence of the exponents on the initial condition \cite{Karl2016}, see also \cite{Johnstone2019a.Science.364.1267} for related experiments.
Intuitively, this is due to the possible existence of, e.g.,~different configurations of linear or nonlinear, such as topological, excitations dominating the dynamics.
Nevertheless, to be physically relevant, a range of initial conditions must lead to the same universal dynamics (see Fig.~\ref{fig0}).
Thus, it is sensible to define, for every NTFP, an associated basin encompassing initial states, from which the system can evolve into the vicinity of that fixed point \cite{Berges2014_attractive_basin}.

In this work, we experimentally confirm the existence of two distinct NTFPs in a system with the same microscopic Hamiltonian. We achieve this by generating initial conditions (ICs) belonging to two different associated basins.
Both ICs are obtained by quenching to the same parameter regime. Prior to the quench, one is prepared in the ground state of another phase, while for the other one we additionally generate localized excitations, employing local spin control.
For the two scaling evolutions we find distinct sets of universal exponents and the corresponding universal scaling functions.

  \begin{figure} 
    \centering
    \includegraphics[width=\linewidth]{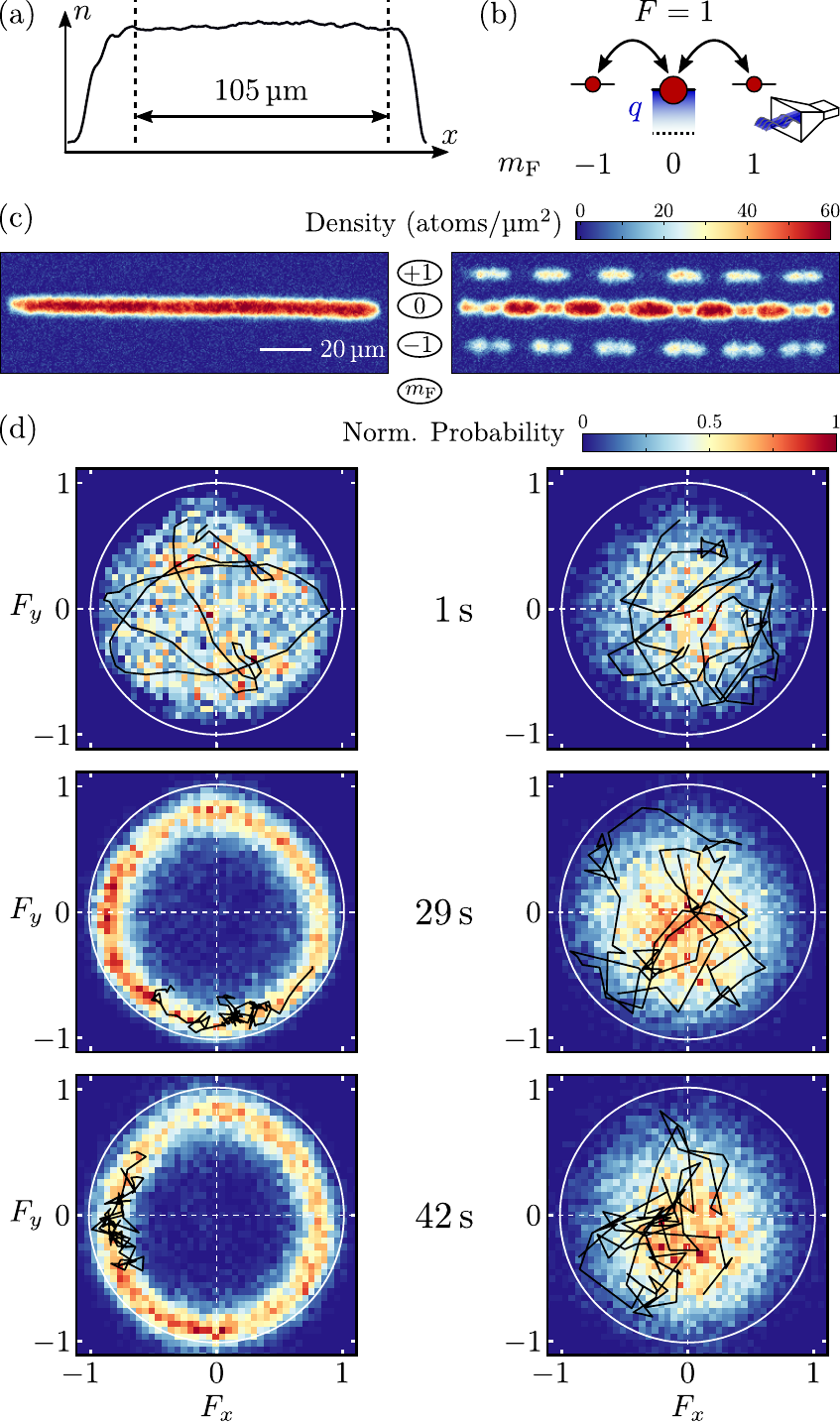}
    \caption{\label{fig1} Time evolution from two different initial conditions (ICs) after quenching the control parameter $q$.
    (a) Quasi-one-dimensional BEC in a box potential (at the start of the evolution at $t=0$) formed by an elongated dipole trap with repulsive walls.
    (b) The relevant energy difference $q$ between the $\mf=0$ and $\mf=\pm1$ levels is controlled with off-resonant microwave dressing. This allows tuning spin-changing collisions into resonance, which redistribute population between the hyperfine levels.
    (c) Absorption pictures of the atom clouds after a short Stern-Gerlach pulse for the polar (left) and rotation ICs (right). For the rotation IC, six local spin rotations transfer atoms from $\mf=0$ to $\mf=\pm1$.
    (d) Time evolution of $F_x$-$F_y$-histograms for the polar (left) and rotation ICs (right). The black lines correspond to the spatial profiles of a representative single realization. For the polar IC the system relaxes to an approx.~constant transverse spin length with phase excitations along the ring. In contrast, combined spin-length and phase fluctuations persist for the rotation IC.}
  \end{figure}
  
  In our experiments, we employ a Bose-Einstein condensate of $^{87}$Rb in the $F=1$ hyperfine manifold. The atom cloud is contained in a quasi one-dimensional box consisting of a red-detuned dipole trap with blue-detuned end caps. The latter confine the atoms within the central part of the longitudinal harmonic potential created by the red-detuned beam along the $x$-direction (the resulting density is shown in Fig.~\ref{fig1}(a)). After preparing the different initial conditions, we study their respective dynamics after a quench.
  For brevity, we denote the ICs without and with additional excitations as polar IC and rotation IC, respectively.
  Absorption pictures of these ICs are shown in Fig.~\ref{fig1}(c).
  
  The polar IC is set by a gas with $\sim1.6\cdot10^5$ atoms in the polar state, all of them occupying the $\mf=0$ magnetic sub-state (cf.~Fig.~\ref{fig1}(c)). For the rotation IC, we prepare the polar state with a lower atom number of $\sim4\cdot10^4$ and additionally generate excitations via 6 equally spaced local spin rotations. Each rotation angle is chosen greater than $\pi/2$, which, in the polar phase, leads to the generation of vector solitons \cite{lannig_collisions_2020}.
  The lower atom number is chosen to ensure a sufficiently long lifetime of excitations.
  
  Following the preparation of each IC, the parameter $q$, denoting the energy offset between $\mf=0$ and $\mf=\pm1$ (see Fig.~\ref{fig1}(b) for the internal level structure and \cite{sm} for the microscopic Hamiltonian), is quenched to a value within the easy-plane phase \cite{kawaguchi_spinor_2012,stamper-kurn_spinor_2013}, by instantaneously applying off-resonant microwave dressing \cite{Gerbier2006_mw_dressing}. The quench tunes spin-changing collisions into resonance, which redistribute atoms between the magnetic levels for both ICs (cf.~Fig.~\ref{fig1}(b)) and lead to a build-up of transverse spin $F_\perp=F_x+\mathrm{i}F_y$, with only small excitations arising in $F_z$ (see Fig.~S1 in \cite{sm}).
  To access the order-parameter field $F_\perp$, we simultaneously measure orthogonal spin projections transverse to the magnetic offset field with spatial resolution along the $x$-axis \cite{kunkel_simultaneous_2019}.
  
 \begin{figure*} 
    \centering
    \includegraphics[width=\linewidth]{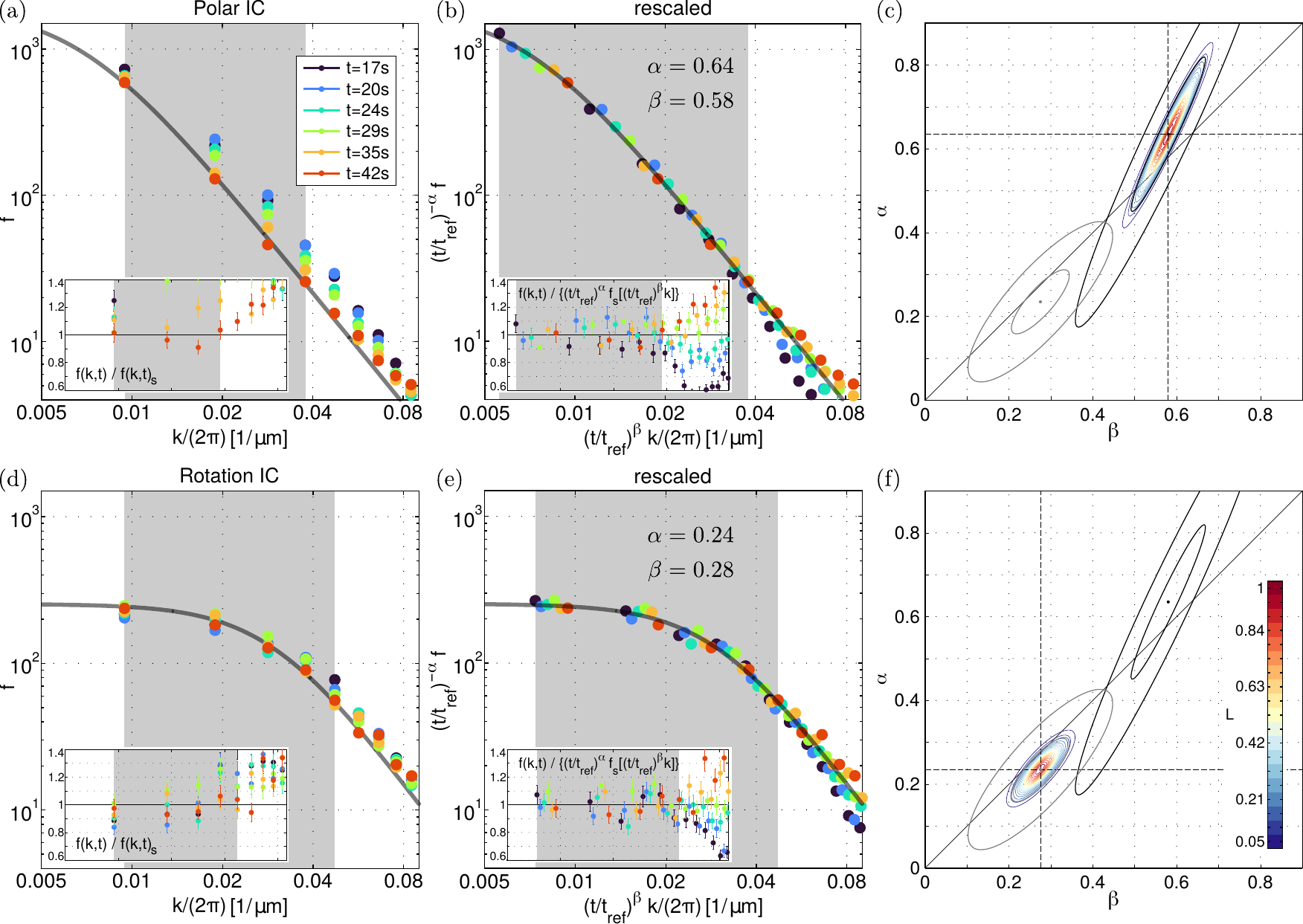}
    \caption{\label{fig2} Scaling evolution of the transverse-spin structure factor $f$ for the polar (a-c) and rotation ICs (d-f).
    Unscaled (a,d) and scaled (b,e) $F_\perp$ power spectra as a function of $k=2\pi/\lambda$ with $\lambda$ the wavelength and corresponding residuals (insets) with respect to the scaling function $f_\text{s}(k)$ at reference time $t_\text{r}=\SI{42}{s}$ are displayed over the time range which shows self-similar evolution. The scaling exponents are extracted via a $\chi^2$ minimization with respect to the function $f_\text{s}$, scaled according to Eq.~(\ref{eq:scaling}) (see \cite{sm} for details). For this analysis, only the points in the gray shaded areas are used, where the spectra have the same shape. All error bars indicate 1 s.d.~of the mean; where no error bars are visible they are smaller than the plot marker.
    This leads to the likelihood function $L\propto\exp(-\chi^2/2)$ (normalized to 1 at the maximum) of the residuals $\chi^2$ shown in the colored contour plot in (c,f). The black and gray ellipses indicate the $2\sigma$ and $5\sigma$ statistical error ranges of the extracted exponents. The optimal scaling exponents given in the text are extracted at the position of the minimal residuals
    (marked by dashed lines and dots).}
  \end{figure*}

  We observe that, after around \SI{17}{s}, the system reaches a regime where long-wavelength correlations only change slowly in time (cf.~Fig.~S2 in \cite{sm}), however, excitations in the transverse spin plane differ drastically for the distinct ICs. For the polar IC the ensuing dynamics after the quench causes the transverse spin $F_\perp$ to settle to the bottom of the corresponding Mexican-hat mean-field potential in the transverse-spin observables, with Goldstone excitations along the bottom in angular direction (see the left col.~of Fig.~\ref{fig1}(d)). Contrary to this, when starting from the rotation IC, large fluctuations filling the transverse plane build up and persist during the subsequent evolution (right col.~of Fig.~\ref{fig1}(d)). We estimate that, for both ICs, the energy input by the quench leaves the system well below the critical temperature for which the easy-plane phase vanishes \cite{Phuc2011_thermal_spinor,sm}. Thus, we conclude that the expected thermal state for these energies has the same properties as a thermal state in the easy-plane phase \cite{Prufer2022_thermalization}; that is, one would expect finding a ring structure in the $F_x$-$F_y$-histograms.
  
  For both initial conditions we infer spatial dynamics which shows scaling in space and time with distinct parameters.
  To investigate this phenomenon, we evaluate the time-evolving structure factor $f(k,t)=\langle\vert \text{dFT}[F_\perp(x,t)](k)\vert^2\rangle$ from the transverse-spin momentum spectra, cf.~Fig.~\ref{fig2}(a,d). Here, $\text{dFT}[\cdot]$ denotes the discrete Fourier transform and $\langle\cdots\rangle$ the ensemble average. We analyze the evolution of $f(k,t)$ with respect to spatio-temporal scaling of the form \cite{orioli_universal_2015,schmied_non-thermal_2019}
  \begin{equation} \label{eq:scaling}
      f(k,t)=\left({t}/{t_\text{r}}\right)^\alpha f_\text{s}\left(\left[t/t_\text{r}\right]^\beta k\right)\,,
  \end{equation}
  where $\alpha$ and $\beta$ denote the amplitude and momentum scaling exponents, $f_\text{s}$ represents a time-independent scaling function and $t_\text{r}$ is a reference time within the scaling regime.

  After an initial period of $\sim\SI{17}{s}$, the spectra $f$ have approached the form of the scaling function $f_\text{s}$. Thereafter, we observe, independently for both ICs, a self-similar shift towards lower momenta that is in accordance with Eq.~(\ref{eq:scaling}) (see Figs.~\ref{fig2}(a,d)).
  Figs.~\ref{fig2}(b,e) show the respective rescaled spectra. These coincide well with scaling functions $f_\text{s}(k)\propto1/[1+(k/k_\text{s})^\zeta]$ (solid gray lines in Figs.~\ref{fig2}(b,e)), with inverse-length scale $k_\text{s}=2\pi/\lambda_\text{s}$ (for $t_\text{r}=\SI{42}{s}$ we find $\lambda_\text{s}=\SI[parse-numbers=false]{(147 \pm 13_{\,\text{stat}})}{\micro\meter}$ for the polar IC and $\lambda_\text{s}=\SI[parse-numbers=false]{(33.5 \pm 1.3_{\,\text{stat}})}{\micro\meter}$ for the rotation IC). This matches the previous observations in a harmonic trap \cite{prufer_observation_2018}. In the re-scaling analysis for the polar IC we find the values
  \begin{align*}
    \alpha
    &=\num[parse-numbers=false]{0.64 \pm 0.09_{\,\text{stat}} \pm 0.50_{\,\text{sys}}}\,, 
    &
    & \\
    \beta
    &=\num[parse-numbers=false]{0.58 \pm 0.04_{\,\text{stat}} \pm 0.26_{\,\text{sys}}}\,,
    &
    \zeta
    &=\num[parse-numbers=false]{2.51 \pm 0.06_{\,\text{stat}}}\,,
  \end{align*}
  while the rotation IC gives rise to scaling with
  \begin{align*}
    \alpha
    &=\num[parse-numbers=false]{0.24 \pm 0.04_{\,\text{stat}} \pm 0.03_{\,\text{sys}}}\,, 
    &
    & \\
    \beta
    &=\num[parse-numbers=false]{0.28 \pm 0.04_{\,\text{stat}} \pm 0.05_{\,\text{sys}}}\,,
    &
    \zeta
    &=\num[parse-numbers=false]{2.87 \pm 0.18_{\,\text{stat}}}\,.
  \end{align*}
  Here, we have chosen $t_\text{r}=\SI{42}{s}$ for the extraction of the scaling functions. The statistical error is extracted from the root-mean-square width of the marginal likelihood distributions shown in Fig.~\ref{fig2}(c,f) and the systematic errors are estimated from the variability of the exponents when varying the momentum cutoff (see \cite{sm} for details).
  The large systematic error for the exponent $\alpha$ in the polar IC is a result of the finite system size, which obstructs the observation of a clear infrared plateau.
  Nevertheless, under the assumption of the transport of conserved $F_\perp$ excitations \cite{Mikheev2019_leeft,pruefer_extraction_2020}, implying $\alpha=\beta$, we obtain $\beta=\num[parse-numbers=false]{0.54 \pm 0.02_{\,\text{stat}} \pm 0.05_{\,\text{sys}}}$ for the polar IC and $\beta=\num[parse-numbers=false]{0.28 \pm 0.03_{\,\text{stat}} \pm 0.04_{\,\text{sys}}}$ for the rotation IC.
  
  For the extraction of the scaling exponents and function parameters we apply a multi-step $\chi^2$ optimization procedure, described in more detail in \cite{sm}. First, approximate scaling exponents are extracted by minimizing the squared deviations between measured spectra with respect to the scaling hypothesis (\ref{eq:scaling}). Then, the parameters of the scaling function are obtained from a fit to all rescaled data. Finally, these parameters enter into a functional scaling model, which is used to define the squared deviations $\chi^2=\sum_{k,t}[f_{k,t}-f(k,t)]^2/\sigma_{k,t}^2$ of the unscaled measured values $f_{k,t}$ from the scaling prediction $f(k,t)$, relative to the measurement uncertainties $\sigma_{k,t}$. The resulting likelihood $L\propto\exp[-\chi^2(\alpha,\beta)/2]$ is shown in Figs.~\ref{fig2}(c,f), which is used to obtain the scaling exponents with statistical errors as stated above.
  We select the momentum scaling regime (gray shaded area in Fig.~\ref{fig2}) from the lowest $k>0$ up to $k_\text{max}$, where the shape of the measured spectra starts to differ for different times. The insets in Fig.~\ref{fig2} depict the residuals of the spectra, obtained by dividing the measured data by the scaling function.

\begin{figure} 
    \centering
    \includegraphics[width=\linewidth]{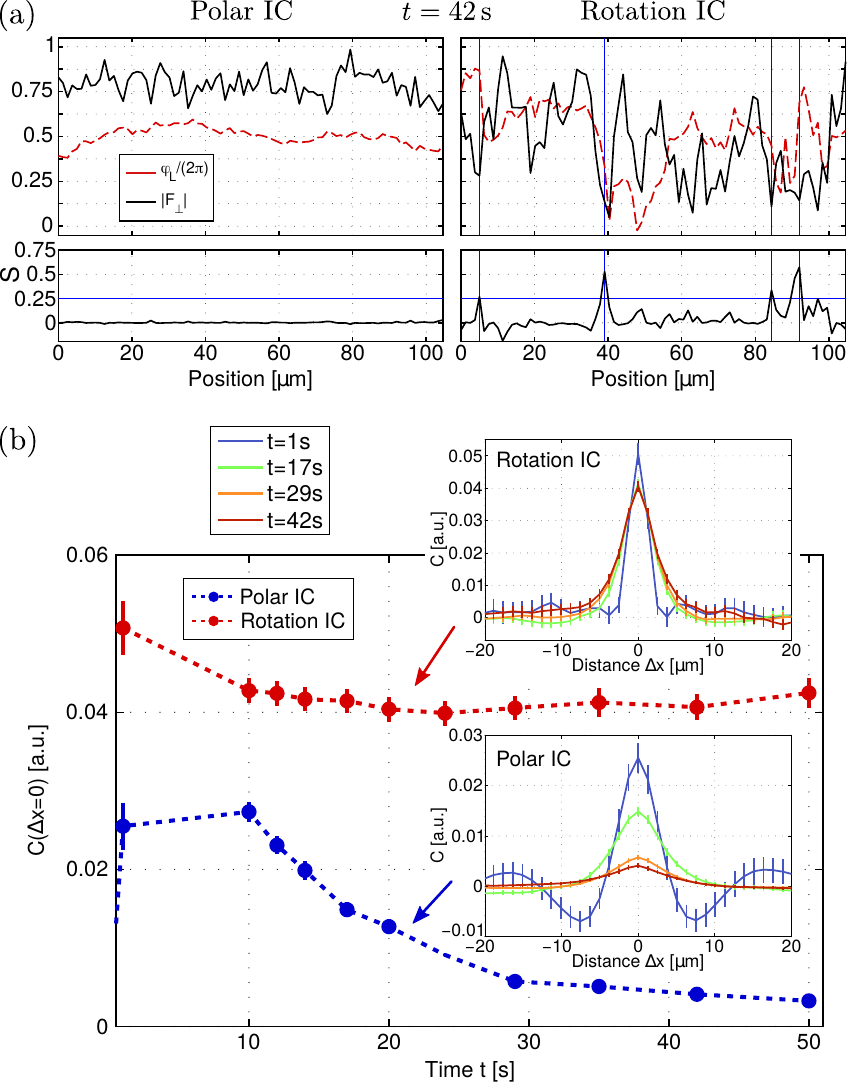}
    \caption{\label{fig3} Spatial structure of the spin excitations in the system.
      (a) Transverse-spin length $\lvert F_\perp\rvert$ (upper plot: solid black line) and Larmor phase $\varphi_\text{L}$ (upper plot: dashed red line) extracted from the same single realization at $t=\SI{42}{s}$ shown as black line in Fig.~\ref{fig1}. The lower plot shows the argument $S(x)=\Delta\vert F_\perp(x)\vert\cdot\abs{\nabla\phil(x)}$ of the correlator given in Eq.~(\ref{eq:length_phase_correlator}) for the single realizations. All positions with large correlations between spin length reduction and phase gradient with $S(x)>0.25$ (horizontal blue line) are marked with vertical blue lines.
      (b) Evolution of correlator amplitude $\mathscr{C}(\Delta x=0,t)$ and spatial profiles (insets) of the full cross correlator $\mathscr{C}$. This shows the larger abundance and persistence of excitations for the rotation IC as compared to the polar IC, which shows a decay.}
  \end{figure}

  To gain more insight into the difference between the two scaling scenarios we evaluate the space-resolved profiles of the transverse spin $F_\perp(x)=\vert F_\perp(x)\vert e^{\mathrm{i}\phil(x)}$ of single realizations in Fig.~\ref{fig3}(a). The solid lines show the spin length $\vert F_\perp(x)\vert$ and the dashed ones the Larmor phase $\phil(x)$ at $t=\SI{29}{s}$. For the polar IC, mostly phase excitations with roughly constant $\vert F_\perp\vert$ are present in the system while, for the rotation IC, strong localized phase-amplitude defects are present (marked by vertical blue lines).

  To characterize these phase-amplitude excitations, we evaluate the spatial cross-correlation function
  \begin{equation} \label{eq:length_phase_correlator}
      \mathscr{C}(\Delta x,t)
      =\sum_x\left\langle\Delta\abs{F_\perp(x,t)}
      \cdot\abs{\nabla\phil(x+\Delta x,t)}\right\rangle
  \end{equation}
  between spin-length variations and gradients of the Larmor phase. Specifically, $\Delta\vert F_\perp\vert(x,t)=\langle{\vert F_\perp\vert}\rangle_x(t)-\vert F_\perp\vert(x,t)$ describes the local deviation of the spin length $\abs{F_\perp}$ from the mean value $\langle{\vert F_\perp\vert}\rangle_x$ taken over all positions and realizations. The time evolution of the local correlator amplitude $\mathscr{C}(\Delta x=0,t)$ in Fig.~\ref{fig3}(b) shows a clear distinction between the decay of excitations for the polar IC as compared to the rotation IC: The spatial correlator profiles show a distinct peak at $\Delta x=0$ in accordance with the structures identified in Fig.~\ref{fig3}(a). While, for the polar IC, the local amplitude $\mathscr{C}(\Delta x=0,t)$ decays over time, it remains approximately unchanged for the rotation IC.
  We find the conservation of the correlator amplitude to be enhanced for lower total atomic densities, while larger densities lead to a decay.

  Numerical simulations for the rotation IC demonstrate that the phase kinks are rather stable, in line with the conservation of $\mathscr{C}(\Delta x=0,t)$, and propagate through the system while interacting with the strongly fluctuating spin \cite{sm}. The corresponding scaling exponents agree with the experimental results.
  This is in contrast to simulations for the polar IC, where a similarly slow scaling with exponents $\alpha=0.27\pm0.06$, $\beta=0.25\pm0.04$ has been observed numerically \cite{schmied_bidirectional_2019,Siovitz2023_rogue_waves}.
  
  We report the measurement of two distinct scaling evolutions characterized by different exponents and scaling functions in a regime with the same microscopic Hamiltonian symmetry of a spinor gas in the easy-plane phase.
  Our findings show the existence of distinct associated basins of initial conditions, that is, the sets of states which evolve towards their respective NTFP.
  Thus, for a classification of universal phenomena far from equilibrium, not only the scaling properties in the vicinity of NTFPs are important but also the structure of their associated basins.
  This hints towards the necessity to include the initial condition for identifying universality classes of far-from-equilibrium dynamics; this is still an unsettled question.

  \begin{acknowledgments}
     The authors thank S.~Erne for useful comments on error estimation and K. Boguslavski, P.~Heinen, P.G.~Kevrekidis, A.N.~Mikheev, and C.M.~Schmied for discussions and collaboration on related topics.
     They acknowledge support by the Deutsche For\-schungs\-gemeinschaft (DFG, German Research Foundation) through SFB 1225 ISOQUANT -- 273811115, and GA677/10-1, as well as under Germany's Excellence Strategy -- EXC-2181/1 -- 390900948 (the Heidelberg STRUCTURES Excellence Cluster),
    and by the state of Baden-W{\"u}rttemberg through bwHPC and DFG through grants INST 35/1134-1 FUGG, INST 35/1503-1 FUGG, INST 35/1597-1 FUGG, and 40/575-1 FUGG.

  \end{acknowledgments}
  

%

\clearpage
\begin{appendix}
\onecolumngrid

\setcounter{equation}{0}
\renewcommand\theequation{S\arabic{equation}}
 \setcounter{table}{0}
 \renewcommand{\thetable}{S\Roman{table}}
\setcounter{figure}{0}
\renewcommand{\thefigure}{S\arabic{figure}}

\section{Supplemental material}
  \subsection{Experimental System}
    \noindent\textbf{Trapping Potential}\\
    The dipole trap in the experiment is formed by a focused \SI{1030}{nm} laser beam, leading to an approximately harmonic confinement with trap frequencies of $(\omega_\parallel,\omega_\perp)=2\pi\times(1.6,160)\,\si{Hz}$, in conjunction with two blue-detuned beams at \SI{760}{nm}. The latter implement walls separated by $\sim\SI{130}{\micro\meter}$, which lead to an approximately homogeneous system with a rapid decrease of the total density within $\sim\SI{10}{\micro\meter}$ at the edges.

    \vspace{2em}
    \noindent\textbf{Microscopic Hamiltonian}\\
    The Hamiltonian of the system is described by \cite{kawaguchi_spinor_2012,stamper-kurn_spinor_2013}
    \begin{equation}
        \hat{H} = \hat{H}_0 + \int dV \biggl[\,: \frac{c_0}{2}\hat{n}^2 + \frac{c_1}{2}\left(\hat{F}_x^2+\hat{F}_y^2+\hat{F}_z^2\right) : + p\left(\hat{n}_{+1}-\hat{n}_{-1}\right) + q\left(\hat{n}_{+1}+\hat{n}_{-1}\right) \biggr]\,,
    \end{equation}
    where $\hat{H}_0$ describes the kinetic energy and trapping potential and $\hat{n}=\sum_j\hat{n}_j$, $\hat{n}_j=\hat{\psi}_j^\dagger\hat{\psi}_j$, with $\hat{\psi}_j^{(\dagger)}$ being the bosonic field annihilation (creation) operator in the magnetic substate $j\in\{0,\pm1\}$, $F_x=[\hat{\psi}_0^\dagger(\hat{\psi}_{+1}+\hat{\psi}_{-1})+\text{h.c.}]/\sqrt{2}$, $\hat{F}_y=[i\hat{\psi}_0^\dagger(\hat{\psi}_{+1}-\hat{\psi}_{-1})+\text{h.c.}]/\sqrt{2}$, $\hat{F}_z=\hat{n}_{+1}-\hat{n}_{-1}$ describe the spin field operators, $:\,:$ denotes normal ordering, and $p$ and $q$ label the linear and second-order Zeeman shifts, respectively. For $^{87}$Rb in the $F=1$ hyperfine manifold the spin interactions are ferromagnetic ($c_1<0$), such that spin-changing collisions (SCCs) of the form $\hat{\psi}_{+1}^\dagger\hat{\psi}_{-1}^\dagger\hat{\psi}_0\hat{\psi}_0+\text{h.c.}$, contained in the terms $\hat{F}_x^2+\hat{F}_y^2$ of the Hamiltonian, lead to the build-up of the transverse spin, as shown in the left column of Fig.~\ref{fig1}(d).

\begin{figure}[b]
      \centering
      \includegraphics[width=0.8\linewidth]{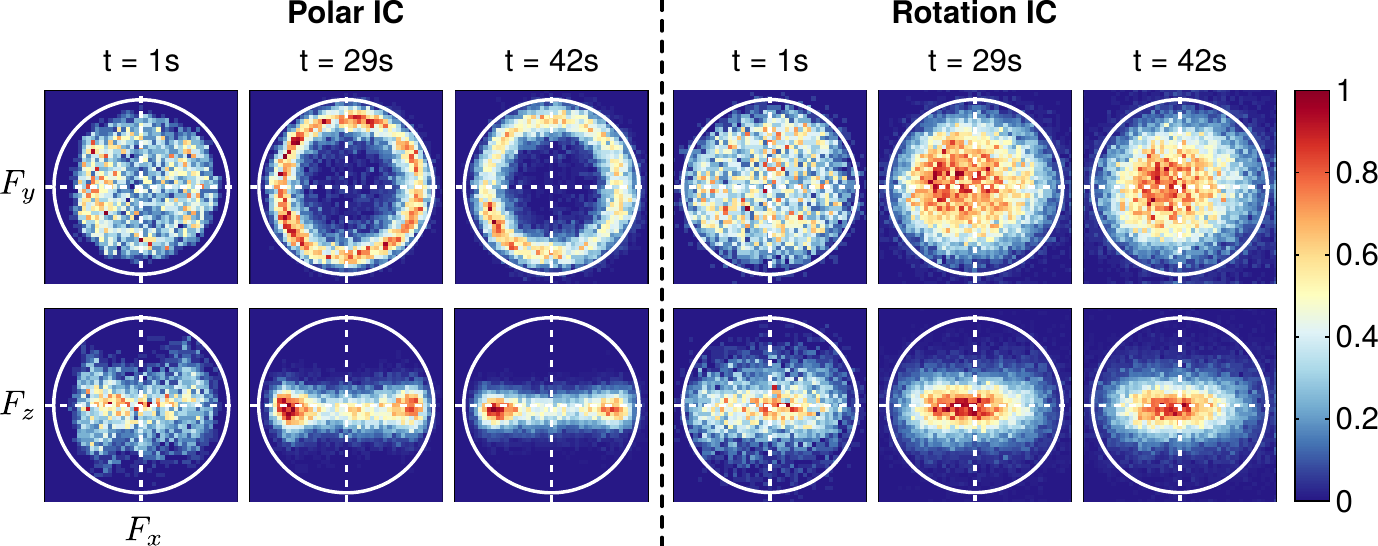}
      \caption{\label{sfig1} Histograms of the $F_x$-$F_y$ and $F_x$-$F_z$ cross sections for a simultaneous measurement of all three spin projections during the time evolution. Throughout the scaling regime ($t=\SIrange{17}{42}{s}$) the $F_z$ fluctuations stay small for both ICs.}
    \end{figure}

    \vspace{2em}
    \noindent\textbf{Magnetic Offset Field and Control of $\bm{q}$}\\
    During the experiment the atom cloud is subject to a homogeneous magnetic offset field of $B=\SI{0.894}{G}$, leading to a linear and quadratic Zeeman splitting of $p=h\times\SI{626}{kHz}$ and $q_\text{B}=h\times\SI{58}{Hz}$, respectively. The spin dynamics is controlled by varying the energy detuning $q=q_\text{B}+q_\text{MW}$ by applying off-resonant microwave dressing \cite{Gerbier2006_mw_dressing} at a Rabi frequency $\Omega_\text{MW}\approx2\pi\times\SI{7.2}{kHz}$ and blue-detuned by $\delta_\text{MW}=2\pi\times\SI{246.256}{kHz}$ (polar IC) or $\delta_\text{MW}=2\pi\times\SI{248.645}{kHz}$ (rotation IC) to the $(F=1,m_\text{F}=0)\longleftrightarrow(F=2,m_\text{F}=0)$ transition. This corresponds to a difference in $q_\text{MW}$ of approx.~$h\times\SI{0.5}{Hz}$. The quench of $q$ at $t=0$ is implemented by instantaneously switching on the microwave power. The detuning $\delta_\text{MW}$ for the polar IC is chosen such that the transverse spin length $\vert F_\perp\vert$ after $t=\SI{30}{s}$ becomes maximal at a value of 0.73. Also the $q$ value used for the rotation IC is close to the maximum of $\vert F_\perp\vert$ in this measurement for the corresponding density without local rotations. For both initial conditions we confirm to stay in the easy-plane phase throughout the dynamical evolution by observing only small excitations in $F_z$ (see Fig.~\ref{sfig1}).
    
    Magnetic field gradients are precisely compensated by tuning $q$ to the left half of the SCC resonance, where $F_z$ excitations are expected. Here, magnetic field gradients along the cloud lead to a build-up of spatial imbalances in the averaged $m_\text{F}=\pm1$ density profiles at large $t$. The magnetic field gradient is compensated by ensuring that the averaged $m_\text{F}=\pm1$ density profiles are homogeneous at $t=\SI{30}{s}$ after the quench from the polar IC.

    \vspace{2em}
    \noindent\textbf{Atom Loss}\\
    Over the duration of the experiment the atom number decreases and between $t=\SIrange{10}{42}{s}$ the atom number is described well by an exponential decrease with $1/e$ lifetime of $\tau=\SI{37}{s}$ (polar IC) and $\tau=\SI{52}{s}$ (rotation IC). At the start of the scaling regime ($t=\SI{17}{s}$) the one-dimensional densities along axial direction are approx.~\SI{700}{atoms/\micro\meter} (polar IC) and \SI{200}{atoms/\micro\meter} (rotation IC).
    After the initial build-up of transverse spin up to $t=\SI{17}{s}$ the average spin length $\langle\vert F_\perp\vert\rangle_x$ stays approximately constant for both ICs.

    \vspace{2em}
    \noindent\textbf{Local Spin Rotations}\\
    For the rotation IC 6 local spin rotations are successively applied along the cloud within a total duration of $<\SI{1}{ms}$, which is much faster than the spin dynamics (characteristic spin time scale $\gtrsim\SI{600}{ms}$). The central rotation angle of each spin rotation is tuned to be larger than $\pi/2$, such that phase jumps are generated in the $m_\text{F}=0$ component, as described in \cite{lannig_collisions_2020}.
    Due to the additional spin dynamics induced by the quench of $q$ the excitations discussed in \cite{lannig_collisions_2020} are presumably not stable in this system.
    All rotations are performed phase-coherently around the same rotation axis.

    \vspace{2em}
    \noindent\textbf{Transverse Spin Readout}\\
    To extract the transverse spin field $F_\perp=F_x+iF_y$ with spatial resolution a simultaneous readout of both transverse projections is applied in conjunction with Stern-Gerlach (SG) absorption imaging as detailed in \cite{pruefer_extraction_2020}. Additionally, at half the time between the two global $\pi/2$ spin rotations, used to map the transverse observables to the projection direction of the SG gradient, a spin-echo $\pi$-pulse is inserted to mitigate effects of magnetic offset field drifts to the readout. In the analysis of the pictures three adjacent pixels are binned together to a size of $\SI{1.2}{\micro\meter}$, which also corresponds to the resolution of our imaging setup \cite{Kunkel2018_EPR}. In order to exclude edge effects from the varying total density only the central $\sim\SI{105}{\micro\meter}$ of the atom cloud are evaluated.

    \vspace{2em}
    \noindent\textbf{Estimation of the Quench Energy}\\
    A numerical comparison of the energies between the ICs shows that the local spin rotations of the rotation IC induce an initial energy offset of approx.~$k_\text{B}\times\SI{210}{pK}$, where $k_\text{B}$ denotes Boltzmann's constant. The additional energy input from the quench across the polar to easy-plane phase transition is on the order of $n\vert c_1\vert\sim k_\text{B}\times\SI{120}{pK}$ (polar IC) and $n\vert c_1\vert\sim k_\text{B}\times\SI{30}{pK}$ (rotation IC) for the initial densities.
    We estimate the experimental spin-interaction energy $n\vert c_1\vert$ by observing the density-dependence of the spin-changing-collision instability. Zero-temperature Bogoliubov instabilities occur below $q_\text{c}=2n\vert c_1\vert$ for a quench of the polar state.
    From the measured atom numbers we estimate a lower limit of the critical temperature $T_\text{c}$ of Bose-Einstein condensation \cite{Dalfovo1999} and obtain $T_\text{c}\approx\SI{60}{nK}$ for the atom number of the polar IC and $T_\text{c}\approx\SI{40}{nK}$ for the rotation IC at $t=\SI{42}{s}$.
    Previously, we have shown that the system can thermalize to a temperature of $\sim\SI{3}{nK}$ \cite{Prufer2022_thermalization}.
    Therefore, the total temperatures in both systems remain well below the estimated critical temperature of $\sim T_\text{c}/4\gg n\vert c_1\vert$ for the polar to easy-plane phase transition at $q\sim n\vert c_1\vert$ \cite{Phuc2011_thermal_spinor}.
    Hence, the equilibrium state, which is expected to be reached, is, for both ICs, in the easy-plane phase, where $\langle\vert F_\perp\vert\rangle>0$.

  \subsection{Extraction of Scaling Exponents}
    As outlined in the main text the scaling exponents are extracted in two stages. The first model-agnostic approach does not make assumptions about the scaling function but enables a first extraction of the scaling exponents. These values are then used to obtain a re-scaled data set which allows the determination of the scaling function with high fidelity. Based on this a model of the spectra is constructed and it facilitates a standard $\chi^2$ analysis based on the scaling hypothesis together with the scaling function.

\begin{figure}
      \centering
      \includegraphics[width=\linewidth]{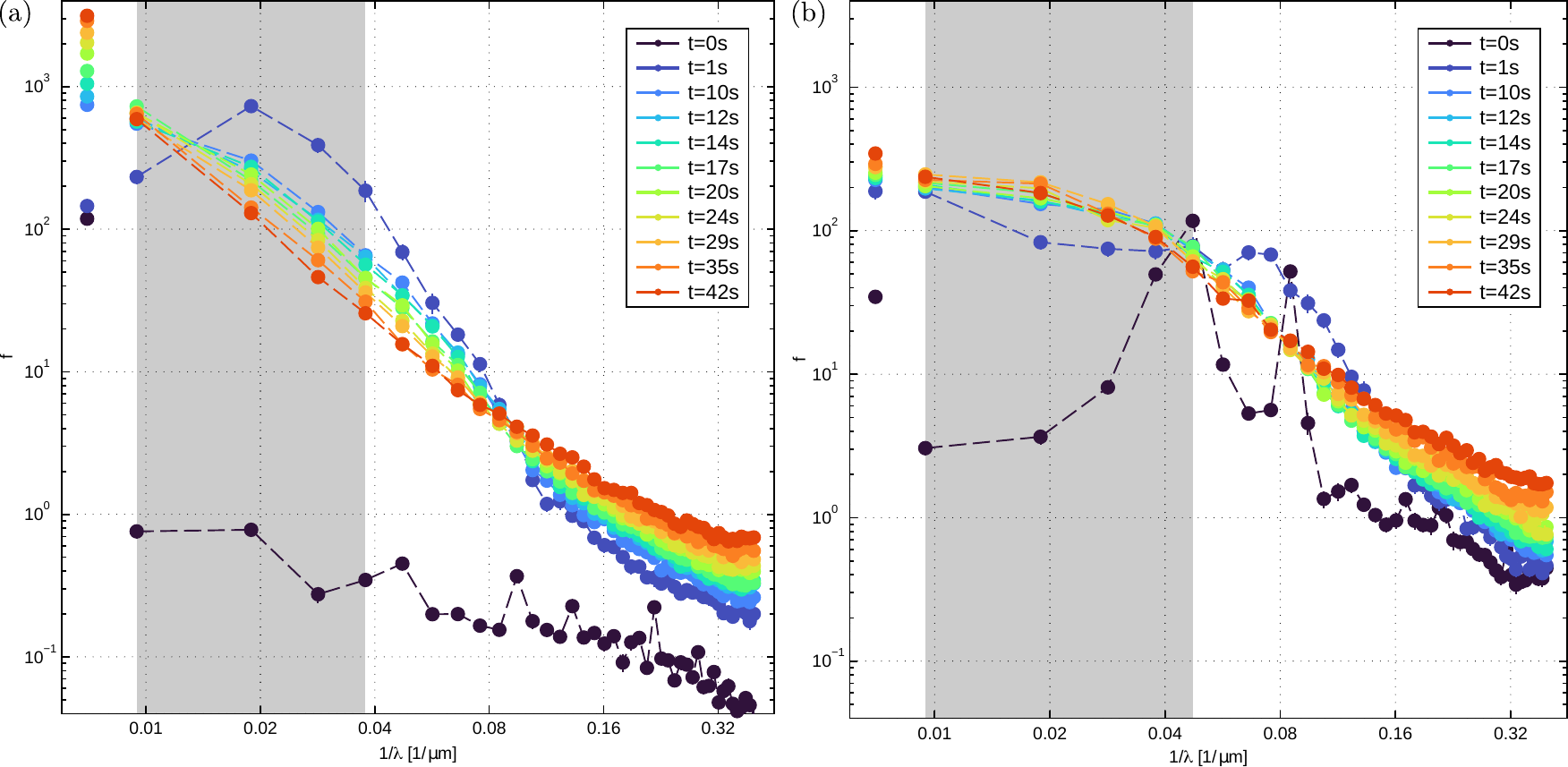}
      \caption{\label{sfig2} Full structure factors for the polar IC (a) and rotation IC (b).
        The data points to the left of the gray shaded region indicate the amplitude of the $k=0$ values.}
    \end{figure}

    \vspace{2em}
    \noindent\textbf{Scaling $\bm{k}$-Range}\\
    Before determining the scaling exponents it is imperative to select the momentum range over which to evaluate re-scaling of the structure factor. This selection process is based on the expected physical process of the transport of excitations to lower $k$ and transport of energy to larger $k$ occurring during the period of self-similar dynamics in the spin-1 system \cite{prufer_observation_2018}. Here, two self-similar regimes have been identified \cite{schmied_bidirectional_2019,glidden_bidirectional_2021}, where exponents are positive for the infra-red (IR) fixed point dynamics at low $k$ and negative for the ultra-violet fixed point at large $k$. Therefore, the $k$-region evaluated for each self-similar process should extend no more than to half the distance between the IR and UV regions marked by the crossing points of the spectra measured at different times to avoid systematic distortion of the extracted scaling exponents. In order not to affect the gradient of the re-scaled spectra at large k by this, we limit the evaluation ranges in $k$-space (gray shaded regions in Figs.~\ref{fig2}(a,d)) to the values
    \begin{align*}
        1/\lambda&=\SIrange{0.009}{0.038}{\per\micro\meter} \;\; \text{(polar IC)}, \\
        1/\lambda&=\SIrange{0.009}{0.047}{\per\micro\meter} \;\; \text{(rotation IC)}.
    \end{align*}
    In this publication we are only evaluating the IR fixed points experimentally. See Fig.~\ref{sfig2} for the full $k$-range of the recorded spectra.
    
    \vspace{2em}
    \noindent\textbf{Error Estimation of the Scaling Exponents}\\
    For the extracted values of the scaling exponents provided in the main text and in Tabs.~\ref{tab:agnostic_results}--\ref{tab:scaling_results_a=b} both statistical and systematic errors are specified.
    After calculating the likelihood functions shown in Figs.~\ref{fig2}(c,f) or \ref{sfig3} the statistical errors are obtained from Gaussian fits to the corresponding marginal likelihood functions. Similarly, the gray ellipses in Figs.~\ref{fig2}(c,f) and \ref{sfig3} are obtained as $n\times\sigma$ intervals extracted from Gaussian fits to the marginal likelihood distributions along rotated axes in $(\alpha,\beta)$-space. The systematic error specifies the change of the exponent values when changing the upper limit of the scaling $k$-range specified above to the neighbouring values.

    \subsubsection{Model-Agnostic Re-Scaling Analysis} \label{sec:model_agnostic_scaling}
\begin{figure}
        \centering
        \includegraphics[width=\linewidth]{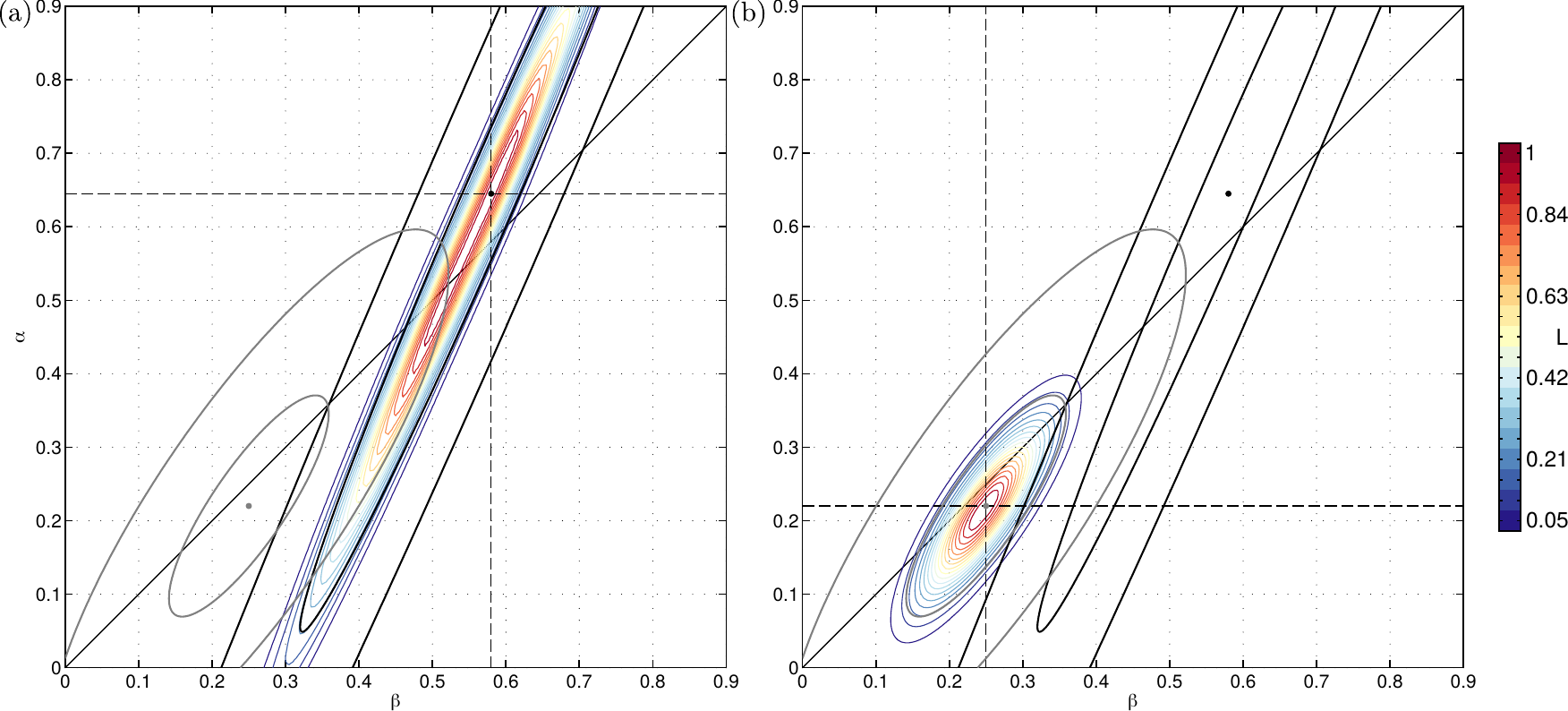}
        \caption{\label{sfig3} Likelihood functions (based on Eqs.~(\ref{eq:model_agnostic_chi2}) and (\ref{eq:likelihood}) for the model-agnostic optimization of the scaling exponents for the polar IC (a) and the rotation IC (b). The black (for polar IC) and gray (for rotation IC) ellipses indicate the $2\sigma$ and $5\sigma$ statistical error ranges.}
      \end{figure}

      The approach outlined here resembles a modified version of the method applied in \cite{erne_universal_2018}.
      For optimizing the scaling exponents without assumptions about the scaling function we use the data measured at an arbitrarily selected reference time $t_\text{r}$ as ''model'' for the remaining measured data. Applying Eq.~(\ref{eq:scaling}) to the data taken at $t_\text{r}$ (where $f_\text{s}$ indicates this measured data) this data is re-scaled to the data at $t\neq t_\text{r}$. This process requires interpolating the reference data at $t_\text{r}$ to the re-scaled $k$ values at $(t/t_\text{r})^\beta k$. For this task we perform a piece-wise power law interpolation (i.e.~a linear interpolation in a bi-logarithmic plot). Based on this approach we define a $\chi^2$ distance measure between the measured data based on the scaling hypothesis (similar to \cite{berges_attractor_2014,erne_universal_2018}) based on the standard errors $\sigma(t,k)$ of the measured mean values of the structure factors $f(t,k)$:
      \begin{equation} \label{eq:model_agnostic_chi2}
          \chi^2_\text{nofct}(\alpha,\beta)=\frac{1}{N_\text{ref}}\sum_{t_\text{r}}\sum_{t,k}\frac{\left\{(t/t_\text{r})^{-\alpha} f[t,(t/t_\text{r})^\beta k]-f(t_\text{r},k)\right\}^2}{\left\{(t/t_\text{r})^{-\alpha}\sigma[t,(t/t_\text{r})^\beta k]\right\}^2+\sigma(t_\text{r},k)^2}\,.
      \end{equation}
      Note that experimental data is described by the ''function'' notation, e.g.~$f(t,k)$ for improved readability (main text: $f_{t,k}$).
      To remove a possible bias from the particular choice of $t_\text{r}$ we average the $\chi^2$ measure over all $N_\text{ref}=6$ reference times (first sum in Eq.~(\ref{eq:model_agnostic_chi2}) in the scaling regime (times $t$ shown in Fig.~\ref{fig2}(a,b,d,e)). Fig.~\ref{sfig3} shows the corresponding normalized likelihood function
      \begin{equation} \label{eq:likelihood}
          L(\alpha,\beta)\propto\exp\left[-\chi^2(\alpha,\beta)/2\right]
      \end{equation}
      with $\chi^2=\chi^2_\text{nofct}$.
      Its minimization over $\alpha$ and $\beta$ yields the optimal exponents given in Tab.~\ref{tab:agnostic_results}. The table also includes the average number $d$ of measured data points for all $t\neq t_\text{r}$ which still lie in the scaling $k$-range after re-scaling and interpolation.
      
      \begin{table}
        \centering
        \setlength{\extrarowheight}{3pt}
        \begin{tabular}{l|c|c|c|c|c}
             \hspace{2pt}\textbf{Initial Condition}\hspace{2pt} & $\bm{\alpha}$ & $\bm{\beta}$ & \hspace{2pt}\textbf{minimal} $\bm{\chi^2_\text{nofct}}$\hspace{2pt} & \hspace{2pt}\textbf{\# Points} $\bm{\tilde{d}}$ \hspace{2pt} & \hspace{2pt}$\bm{\chi_\text{red}^2=\chi_\text{nofct}^2/\nu}$\hspace{2pt} \\[5pt]
             \hline
             \hspace{1em} Polar IC     & \hspace{2pt}\num[parse-numbers=false]{0.65 \pm 0.30_{\,\text{stat}} \pm 0.48_{\,\text{sys}}}\hspace{2pt} &
                                         \hspace{2pt}\num[parse-numbers=false]{0.58 \pm 0.13_{\,\text{stat}} \pm 0.24_{\,\text{sys}}}\hspace{2pt} &
                                         \hspace{2pt}\num[parse-numbers=false]{25.4}\hspace{2pt} &
                                         135 &
                                         \hspace{2pt}\num[parse-numbers=false]{1.24}\hspace{2pt} \\
             \hspace{1em} Rotation IC  & \num[parse-numbers=false]{0.22 \pm 0.08_{\,\text{stat}} \pm 0.02_{\,\text{sys}}} &
                                         \num[parse-numbers=false]{0.25 \pm 0.05_{\,\text{stat}} \pm 0.02_{\,\text{sys}}} &
                                         \num[parse-numbers=false]{25.3} &
                                         165 &
                                         \num[parse-numbers=false]{0.99} \\[2pt]
        \end{tabular}
        \caption{\label{tab:agnostic_results}Results of the model-agnostic likelihood maximization for the data given in Fig.~\ref{sfig3}. Here, the number of points $\tilde{d}$ includes data across all $N_\text{ref}$ reference times, such that $d=\tilde{d}/N_\text{ref}$. By accounting for the two scaling exponents, the number of degrees of freedom is estimated to be $\nu=d-2$.}
        \setlength{\extrarowheight}{0pt}
      \end{table}

    \subsubsection{Re-Scaling Analysis with Scaling Function} \label{sec:functional_scaling}
      From the re-scaled data set, obtained with the exponents from Tab.~\ref{tab:agnostic_results}, we extract the parameters of the scaling function $f_\text{s}$, as provided in the main text, from a fit in the scaling $k$-range, weighted with the scaled and interpolated errors $1/[(t/t_\text{r})^{-\alpha}\sigma(t,(t/t_\text{r})^\beta k)]^2$.

      \begin{table}
        \centering
        \setlength{\extrarowheight}{3pt}
        \begin{tabular}{l|c|c|c|c|c}
             \hspace{2pt}\textbf{Initial Condition}\hspace{2pt} & $\bm{\alpha}$ & $\bm{\beta}$ & \hspace{2pt}\textbf{minimal} $\bm{\chi^2}$\hspace{2pt} & \hspace{2pt}\textbf{\# Points} $\bm{d}$ \hspace{2pt} & \hspace{2pt}$\bm{\chi_\text{red}^2=\chi^2/\nu}$\hspace{2pt} \\[3pt]
             \hline
             \hspace{1em} Polar IC     & \hspace{2pt}\num[parse-numbers=false]{0.64 \pm 0.09_{\,\text{stat}} \pm 0.50_{\,\text{sys}}}\hspace{2pt} &
                                         \hspace{2pt}\num[parse-numbers=false]{0.58 \pm 0.04_{\,\text{stat}} \pm 0.26_{\,\text{sys}}}\hspace{2pt} &
                                         \hspace{2pt}\num[parse-numbers=false]{29.5}\hspace{2pt} &
                                         24 &
                                         \num[parse-numbers=false]{1.55} \\
             \hspace{1em} Rotation IC  & \num[parse-numbers=false]{0.24 \pm 0.04_{\,\text{stat}} \pm 0.03_{\,\text{sys}}} &
                                         \num[parse-numbers=false]{0.28 \pm 0.04_{\,\text{stat}} \pm 0.05_{\,\text{sys}}} &
                                         \num[parse-numbers=false]{42.9} &
                                         30 &
                                         \num[parse-numbers=false]{1.72} \\[2pt]
        \end{tabular}
        \caption{\label{tab:scaling_fct_results}Results of the likelihood maximization based on the scaling function and hypothesis as model for the data given in Fig.~\ref{fig2}. The $p$-values are estimated assuming a $\chi^2$ distribution of $\nu=d-5$ degrees of freedom and $\gamma$ specifies the angle of the semi-major axis of the likelihood distribution in Fig.~\ref{fig2}(c,f) with respect to the horizontal axis.}
        \setlength{\extrarowheight}{0pt}
      \end{table}
      Using this scaling function and the scaling hypothesis as model to define $\chi^2$ and applying a maximum likelihood optimization as described in the main text we obtain the values provided in Tab.~\ref{tab:scaling_fct_results}. Because no interpolation is involved in the calculation of $\chi^2$ in this case the total number $d$ of measured data points across all times in the scaling regime stays constant. As the number of degrees of freedom we set $\nu=d-5$ to account for the two re-scaling exponents and amplitude, scale $\lambda_\text{s}$ and power law exponent $\zeta$ of the scaling function as model parameters.

    \subsubsection{Re-Scaling Analysis under the Constraint \texorpdfstring{$\alpha=\beta$}{a=b}}
      The finite size of atom cloud limits the infra-red resolution of the structure factor in the experiment, which induces a large uncertainty in primarily the exponent $\alpha$. From earlier investigations of the polar IC one expects a transport of conserved $F_\perp$ excitations near the NTFP. This implies a conservation of the integral of the structure factor, which fixes $\alpha=\beta$ in a one-dimensional system. Therefore, we are also performing the re-scaling analysis described above under this constraint. Instead of evaluating the two-dimensional distributions in the $\alpha$-$\beta$-plane (cf.~\ref{sfig3}), here only the intersection of the likelihood $L$ along the diagonal is evaluated. The results are given in Tab.~\ref{tab:scaling_results_a=b}.

      \begin{table}
        \centering
        \setlength{\extrarowheight}{3pt}
        \begin{tabular}{l|c|c|c|c}
            \hspace{2pt}\textbf{Initial Condition + Evaluation}\hspace{2pt} & \textbf{Exponent} $\bm{\alpha=\beta}$ & \hspace{2pt}\textbf{minimal} $\bm{\chi^2}$\hspace{2pt} & \hspace{2pt}\textbf{\# Points}\hspace{2pt} & \hspace{2pt}$\bm{\chi_\text{red}^2=\chi^2/\nu}$\hspace{2pt} \\[3pt]
            \hline
            \hspace{1em} Polar IC (model-agnostic) & \hspace{2pt}\num[parse-numbers=false]{0.53 \pm 0.04_{\,\text{stat}} \pm 0.04_{\,\text{sys}}}\hspace{2pt} &
                                                     \hspace{2pt}$\chi^2_\text{nofct}=\num[parse-numbers=false]{25.4}$\hspace{2pt} &
                                                     $\tilde{d}=135$ & 
                                                     \num[parse-numbers=false]{1.24} \\
            \hspace{1em} Polar IC  & \num[parse-numbers=false]{0.54 \pm 0.02_{\,\text{stat}} \pm 0.05_{\,\text{sys}}} &
                                     $\chi^2=\num[parse-numbers=false]{29.0}$ &
                                     $d=24$ &
                                     \num[parse-numbers=false]{1.53} \\[5pt]
            \hspace{1em} Rotation IC (model-agnostic) & \num[parse-numbers=false]{0.24 \pm 0.08_{\,\text{stat}} \pm 0.01_{\,\text{sys}}} &
                                                        $\chi^2_\text{nofct}=\num[parse-numbers=false]{24.6}$ &
                                                        $\tilde{d}=165$ &
                                                        \num[parse-numbers=false]{0.96} \\
            \hspace{1em} Rotation IC  & \num[parse-numbers=false]{0.28 \pm 0.03_{\,\text{stat}} \pm 0.04_{\,\text{sys}}} &
                                        $\chi^2=\num[parse-numbers=false]{44.1}$ &
                                        $d=30$ &
                                        \num[parse-numbers=false]{1.76} \\[2pt]
        \end{tabular}
        \caption{\label{tab:scaling_results_a=b}Results of the likelihood maximization under the constraint $\alpha=\beta$. Refer to Tabs.~\ref{tab:agnostic_results} and \ref{tab:scaling_fct_results} for the definitions of the quantities given here. Due to the constraint on the exponents, the number of degrees of freedom is $\nu=d-4$ for the evaluation based on the scaling function. The fit of the scaling function under the constraint $\alpha=\beta$ results in the same parameters as given in the main text.}
        \setlength{\extrarowheight}{0pt}
      \end{table}

    \subsection{Numerical Simulations}
      We compare the experimental data to Truncated-Wigner (TW) simulations of a homogeneous one-dimensional spinor gas, simulated by solving the spin-1 Gross-Pitaevskii equations subject to periodic boundary conditions, similar to the simulations performed in \cite{Siovitz2023_rogue_waves,schmied_bidirectional_2019}. Local spin rotations with an angle of $2\pi/3$ are imprinted onto the initial condensate to roughly result in the same density of excitations as in the experiment with respect to the spin-healing length $\xi_\text{s}=\hbar/\sqrt{2mn|c_1|}$, where $\hbar$ denotes the reduced Planck constant, $m$ the atomic mass, $n$ the density and $c_1$ the spin-interaction constant. To initiate dynamics the second-order shift is quenched to $q=0.9n|c_1|$. The dynamical evolution is then characterized in terms of the spin-changing-collision time scale $t_\text{s}=h/(n|c_1|)$. We note that the simulations are performed with a much larger atom number $\sim 3 \cdot 10^6$ compared to the experiment, resulting in a physical density of $n=\SI{13600}{\per\micro\meter}$.

\begin{figure}
      \centering
      \includegraphics[width=0.75\linewidth]{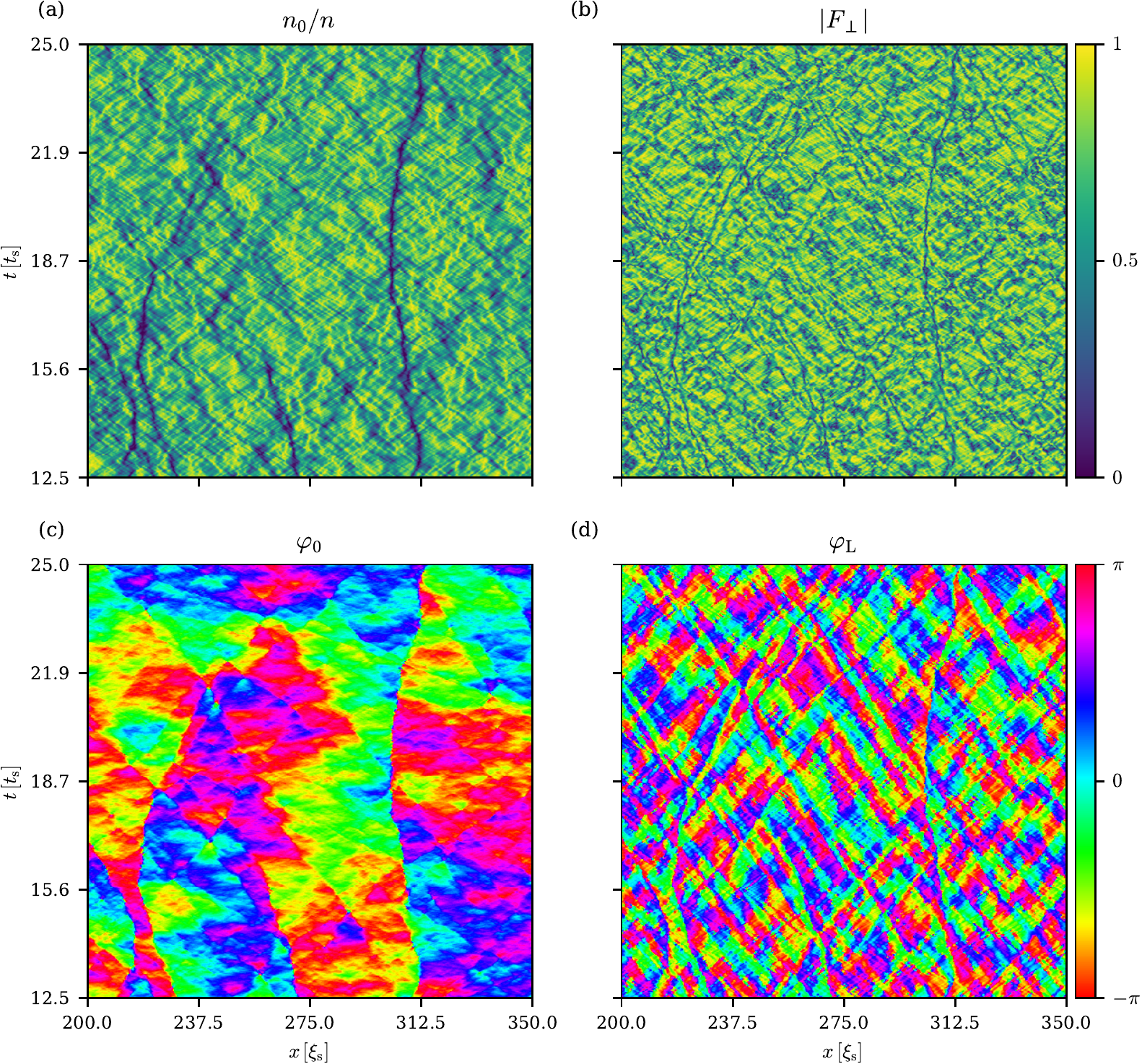}
      \caption{\label{sfig4} Numerical simulation of the time evolution after a quench for the rotation IC in the observables (a) $n_0/n$,  (b) $\vert F_\perp\vert$,  (c) the phase $\varphi_0$ of the $\mf=0$ field and (d) $\phil$. The data shows a classically propagated single realization of the full TW simulation. Around $x\approx\SI{312}{\xi_{\text{s}}}$ long-lived density depletion in the $\mf=0$ component is observed to propagate through the condensate, accompanied by a phase jump of $\approx \pi$. This excitation is also seen in the transverse spin degree of freedom.}
    \end{figure}

      The time evolution of the spin field $F_\perp$, displayed in Fig.~\ref{sfig4}, shows long-lived reductions in length associated with a jump of the Larmor phase $\phil$.

\begin{figure}
      \centering
      \includegraphics[width=\linewidth]{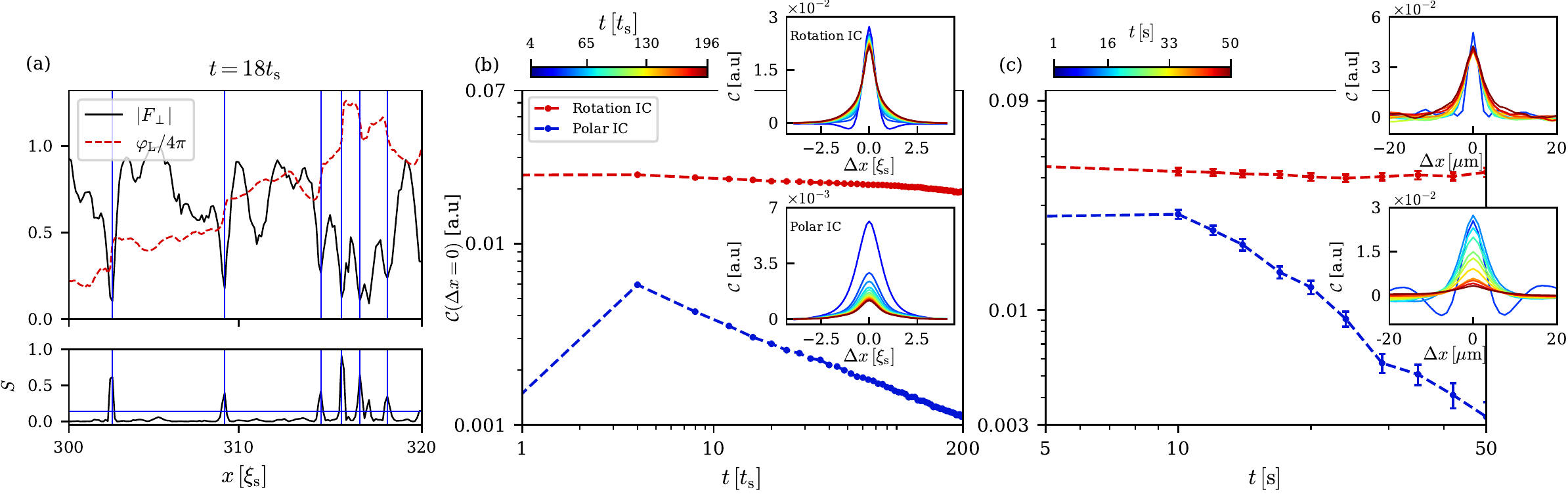}
      \caption{\label{sfig5} Comparison of experimental data with the numerically extracted structure of spin excitations from TW simulations analogous to Fig.~\ref{fig3}.
      (a) An excerpt of a single-time snapshot of the TW run shown in Fig.~\ref{sfig4} at time $t = 18 t_\mathrm{s}$ and $x \in [300,320]\xi_\mathrm{s}$. The lower panel shows the correlator density as defined in Fig.~\ref{fig3}. Events which are larger than a standard deviation of the data are marked by blue lines. (b) Evaluation of the cross-correlator amplitudes $\mathscr{C}(\Delta x =0, t)$ on a double logarithmic scale for the numerical simulation for polar and rotation ICs. The spatial profiles shown in the insets display the same behavior as the experimental data. (c) For comparison, the experimentally extracted correlator amplitudes $\mathscr{C}(\Delta x=0,t)$ of Fig.~\ref{fig3}(b) are shown double-logarithmically, along with the profiles $\mathscr{C}(\Delta x,t)$. Error bars indicate 1 s.d.~of the mean.}
    \end{figure}

      Upon evaluation of the transverse-spin phase-amplitude excitations, the numerical simulations show qualitative agreement with the experimental results provided in Fig.~\ref{fig3}. The corresponding numerical data is shown in Fig.~\ref{sfig5}, where the data for the polar IC is the same as presented in \cite{Siovitz2023_rogue_waves}.

\begin{figure}
      \centering
      \includegraphics[width=0.9\linewidth]{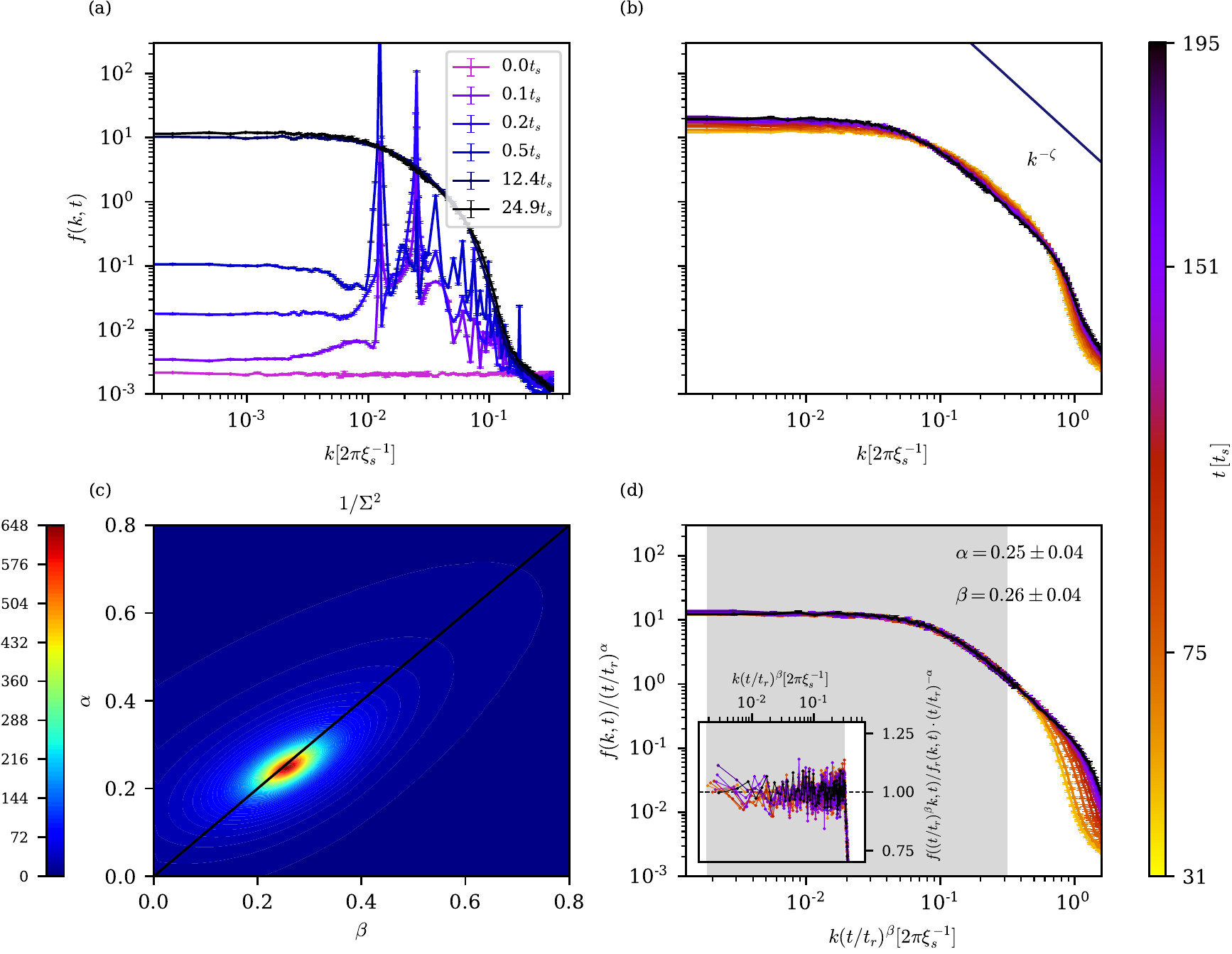}
      \caption{\label{sfig6} Self-similar scaling of the structure factor $f(k,t) = \langle |\mathrm{dFT} \left[| F_\perp (x,t) |\right] (k)|^2\rangle$ for the numerical simulations of the rotation IC.
      (a) Initial condition and short-time dynamics after imprinting an ensemble of 80 rotations onto a condensate of length $L=\SI{220}{\micro\meter}$ and quenching the quadratic Zeeman shift to $q=0.9 n|c_1|$.
      The build-up of structure through instabilities and redistribution of momentum leads to the formation of the universal scaling function.
      (b) Unscaled spectra in the self-similar scaling regime. We extract the power law $\zeta$ via a least-squares fit of the function $f_\mathrm{s}$ and find $\zeta = 2.002 \pm 0.005$.
      (c) Inverse residuals $1/\Sigma^2$ of the rescaling procedure computed according to the prescription given in \cite{orioli_universal_2015}.
      (d) Rescaled spectra of the rotation IC shown in panel (b). The numerical analysis finds scaling exponents $\alpha = 0.25 \pm 0.04$ and $\beta = 0.26 \pm 0.04$. 
      The inset shows the residuals after the rescaling.}
    \end{figure}

      A numerical analysis of the transverse-spin power spectra results in scaling exponents $\alpha = 0.25\pm 0.04$ and $\beta = 0.26\pm 0.04$ (see Fig.~\ref{sfig6}), hence agreeing with experimental results within the error bounds.
      By imprinting local spin rotations, overoccupied modes in the transverse spin are enhanced by Bogoliubov instabilities and redistribute to form a time-independent scaling function $f_\mathrm{s}$ with a large plateau. After around $30t_\mathrm{s}$, the self-similar scaling regime is reached.
      
\end{appendix}
\end{document}